\documentclass[11pt,a4paper]{article}
\usepackage{amsmath}
\usepackage{latexsym}
\usepackage{theorem}
\newtheorem{teorema}{Theorem}[section]
\newtheorem{definicion}[teorema]{Definition}
\include{amslatex}

{\theorembodyfont{\rmfamily}
}
{\theorembodyfont{\rmfamily} }

\numberwithin{equation}{section}

\include{amsmath}

\begin{document}
\begin{title}
{\LARGE {\bf On a covariant version of Caianiello's Model}}
\end{title}
\maketitle
\author{

\begin{center}Ricardo Gallego Torrome\\[3pt]
\paragraph{}
Calle Alicante, 4-6\\[3pt]
46910, Benetusser\\[3pt]
Spain
\end{center}}
\begin{abstract}
Caianiello's  derivation of Quantum Geometry through an isometric
embedding of the spacetime $({\bf M},\tilde{g})$ in the
pseudo-Riemannian structure $({\bf T^*M},g^*_{AB})$ is reconsidered.
In the new derivation, a non-linear connection and the bundle 
formalism induce a Lorentzian-type structure
in the $4$-dimensional manifold ${\bf M}$ that is covariant 
under arbitrary local coordinate transformations in {\bf M}. If
models with maximal acceleration are required to be non-trivial,
gravity should be supplied with other interactions in a unification
framework.
\end{abstract}
\section{Introduction}
The maximal proper acceleration of a massive particle has been 
introduced by Caianiello as a consequence of its re-interpretation of Quantum Mechanics in 
the contest of Information Theory and System Theory ([1]). In
Caianiello's theory, the value of the maximal acceleration is given
by the relation:
\begin{equation}
A_{max}:= \frac{2 m c^3 }{\hbar}.
\end{equation}
This value was obtained considering the evolution of a free particle in a flat, torsion-free phase-space 
and constitutes a notable element of his theory. Relevant too, the
different interpretations can in principle be checked experimentally
and their refutation can be an striking test of their supporting
frameworks.

There are two different kinds of interpretations of the formula $(1.1)$. In the first one, $m$ is 
the rest mass of the particle being accelerated. In the second
interpretation, $m$ is an universal mass scale,
 typically of the same order than Planck's mass $M_p $. Indeed, for $m\sim M_p$ the order 
of the maximal acceleration coincides with the value obtained from
string theory ([2]). Another possibility was considered in ref. [7],
where the maximal acceleration has an universal value, corresponding
with $m$ of the order the lightest neutrino's mass. It is important
to note that in all these interpretations, the value of the maximal
acceleration is given in terms of relativistic constants and that it
is invariant under arbitrary local coordinate changes.

We state the index convention used in this note. Indices denoted by
minor and greek letters run from $0$ to $3$, while capital indices
run from $0$ to $7$. If the contrary is not stated, Einstein's
convention should be understood. In Caianiello's Quantum Geometry
the spacetime manifold ${\bf M}$ is $4$-dimensional,
the tangent bundle {\bf TM} 
is $8$-dimensional and the projection $\pi:{\bf TM}\longrightarrow {\bf M}$ induces an 
effective $4$-dimensional geometry different from the original metric geometry of ${\bf 
M}$. This can be achieved through an embedding procedure ([3]).
As result, the metric of 
the space-time ${\bf M}$ is modified from
\begin{displaymath}
ds^2 _0 =g_{\mu \nu} d x^{\mu} d x^{\nu}, \quad \mu ,\nu =0,1,2,3
\end{displaymath}
to the new line element
\begin{equation}
ds^2 =\Big(1+ \frac{  \ddot{x}^{\sigma}(s_0 ) \ddot{x}_{\sigma}(s_0
)}{A^2 _{max}}\Big)\, ds^2 _0=\lambda(\ddot{x}(s_0))ds^2 _0,\quad
\sigma =0,1,2,3 .
\end{equation}
$g_{\mu \nu} (x)$ is the initial Lorentzian spacetime metric at the point $x\in {\bf M}$ and 
$\ddot{x}^{\mu}$ are the components of the acceleration at this
point, $\ddot{x}^{\mu}(s_0)=\frac{d^2 x^{\mu}}{ds^2 _0}$. When it is 
possible to invert the equation $\dot{x}^{\mu}(s_ 0)=y^{\mu }(s_0 )$, the immersion 
procedure is an embedding and the metric $(1.2)$ lives in {\bf M},
because the factor $\lambda(\ddot{x}(s_0))\rightarrow
\tilde{\lambda}(x(s_0))$ lives in ${\bf M}$. However,
 $ \tilde{\lambda} (x(s_0))$ is not an invariant factor and therefore the line element (1.2) is not
invariant under arbitrary
coordinate transformations of the 
spacetime manifold ${\bf M}$.

The aim of this note is to provide a solution to this covariance
problem using the minimal geometric content and standard methods
from Differential Geometry. In this sense, the paper is a minimal
extension of the initial model, compared with other attempts to
describe the geometry of maximal acceleration ([16]).  In addition
to the original embedding procedure suggested by Caianiello et al.
([3]), we introduce an alternative approach, where a higher order
non-degenerate Finsler-Lagrange structure appears. These type of
structures were extensively studied by Miron's school ([13], [20]),
at least in the positive definite case.

The present paper is organized in the following way. First, we
review in {\it section 2}
the deduction of {\it equation} $(1.2)$ in usual Quantum Geometry. Then, after the 
introduction of the notion of non-linear connection
in {\it section 3}, we understand the covariance problem 
and we show how using the non-linear connection is possible to solve
it. In {\it section 4} we re-derive the immersion interpretation of
$(1.2)$ but using the correct formalism introduced in {\it section
3}. We also discuss the usefulness of the standard embedding
procedure and we describe an alternative
interpretation of the original formalism. In 
order to provide a general framework for the new formulation, we
briefly
introduce the relation of maximal acceleration 
with Finslerian Deterministic Systems. Finally, a discussion of some implications of the new 
formulation and its connection with the old one is also presented in 
{\it section 5}.
\section{Elements of Caianiello's Model}
We review the formal procedure of the derivations of the Quantum
Geometry in Caianiello's  model.
Let us consider the 
$8$-dimensional tangent bundle ${\bf TM}$. It is endowed with a
pseudo-Riemannian metric defined by the expression
\begin{equation}
ds^2 = g_{AB}dX^A dX^B,\quad A,B =0,...,7
\end{equation}
and where the natural coordinates are defined by
\begin{displaymath}
X^A =(x^{\mu},y^{\mu})=\Big (x^{\mu}; \frac{c^2}{A_{max}}\frac{d x^{\mu}}{ds_0 }\Big),\quad 
\mu =0,...,3.
\end{displaymath}
The metric coefficients $g_{AB}$ is given in terms of the space-time
metric $g_{\mu \nu}$ by
\begin{displaymath}
g_{AB}=g_{\mu \nu }\oplus g_{\mu \nu}.
\end{displaymath}
The associated line element in ${\bf TM}$ is expressed as
\begin{equation}
ds^2 =\Big(dx^{\mu }dx^{\nu }+ \frac{c^4 }{A^2 
_{max}}d\dot{x}^{\mu}d\dot{x}^{\nu}\Big)g_{\mu \nu},
\end{equation}
where it was supposed that the set
$\{dx^{\mu},d\dot{x}^{\mu}\}=\{dx^{\mu},dy^{\mu}\}$ is a basis for
the dual frame along a possible trajectory ([3]). The embedding
procedure requires the introduction of a timelike congruence,
associated with the trajectories of particles with fixed positive
mass. Indeed, the original Caianiello's argument introduces an
embedding procedure depending on this congruence, jointly with a
quantum mechanical probability density, in order to reproduce
Quantum Mechanics in the spacetime $({\bf M}, \tilde{g})$ induced
from the phase space geometry $({\bf T^* M},g^*_{ÄB})$, with
$g^*_{AB}=g^{AB}=g^{-1}_{AB}$. Then, the Lorentzian-type structure
that they induce on ${\bf M}$ from the initial metric $(2.1)$
depends on the particular timelike congruence and this is why is not
a properly Lorentzian structure.

Let us consider in some detail the embedding procedure. It can be
understood in a simple way
if the vector field 
${\frac{dx^{\mu}}{d s_0},\, \mu=0,...,3 }$ is given in terms of the coordinates of the 
particle along its physical trajectory. That means that trajectories
are injective curves, which is very likely the case if these
trajectories are in spacetime, where there are no physical loops due
to causality. This was the method used by Caianiello and
co-workers's papers: there is a $one$ to $one$ correspondence
between the values of the parameter $s_0$ and the points $x(s_0)$.
The embedding procedure consist on the inversion $x(s_0 )\rightarrow
s_0 (x)$, where $s_0$ is the length of the physical trajectory,
counting from a fixed point. This produce a non-local dependence on
the effective metric, due to the intrinsic dependence on the whole
trajectory, as discussed before.

In Caianiello's model, an associated metric structure is defined for
the co-tangent bundle ${\bf T^* M}$. Caianiello's model is based on
the fact that the geometry of the
associated pseudo-Riemannian cotangent bundle provides a geometric 
description of Quantum Mechanics where, for instance, the curvature tensor components 
obtained from the above metric are related with Heisenberg's
indeterminacy relations. Maximal acceleration is obtained as a
consequence of this approach to the foundations of Quantum Mechanics
([1]).

Following with the deduction of Caianiello's model, for a particle
of mass $m$, the line element $(2.2)$ is reduced to (1.2). The associated structure has a metric 
components given by
\begin{equation}
\tilde{g}_{\mu \nu}=\Big(1+ \frac{\ddot{x}^{\sigma}\ddot{x}_{\sigma}}{A^2 
_{max}}\Big)g_{\mu \nu},
\end{equation}
which depend on the squared length of the space-time $4$-acceleration $\|\ddot{x}\|^2 
:=g_{\mu \nu}\ddot{x}^{\mu}\ddot{x}^{\nu}$. The term
\begin{displaymath}
h_{\mu \nu}=\frac{\ddot{x}^{\sigma}\ddot{x}_{\sigma}}{A^2
_{max}}g_{\mu \nu}
\end{displaymath}
is called the quantum correction, because it vanishes when $\hbar$
goes to zero.

There is another possibility that is to consider (2.3) as a higher
order Lagrange structure, investigated by the group Miron's school
([13], [20]). A a modern approach to these structures, with
treatment of spinor and other related geometric topics, is compiled
in [12] and also ref. [14] is of interest at this point. In these
references, the mathematical formalism of diverse non-Riemannian
metric structures is exposed and formulated with generality.
\section{The Non-Linear Connection}
In this section we introduce the minimal mathematical
notions and tools that we need in order to 
formulate in a covariant way Caianiello's Quantum Geometry Model.
The references on basic Finsler Geometry that we use are [4] and
[20], and for its higher order generalizations is [20],
adapting definitions and notions to the formulation in the case of 
non-degenerate Finsler and higher order structure. The major changes
with these references consists on using non-homogeneous tensors and
structures, instead of the usual homogeneous of degree zero
expressions found in standard treatments.
\begin{definicion}
A non-degenerate Finsler structure $F$ defined on the
$n$-dimensional manifold ${\bf M}$ is a real function  $F^2:{\bf
TM}\rightarrow [0,\infty [$ such that it is homogeneous of dimension
$2$ in $y$ and it is smooth in the split tangent bundle ${\bf
N}={\bf TM}\setminus \{0\}$ and the hessian matrix
 \begin{equation}
g_{\mu \nu}(x,y): =\frac{1}{2}\frac{{\partial}^2 F^2
(x,y)}{{\partial}y^{\mu} {\partial}y^{\nu} }
\end{equation}
is non-degenerate in ${\bf N}$. The particular case when the
manifold is $4$-dimensional and $g_{\mu \nu}$ have signature
$(+,+,+,-)$ is called Finsler spacetime ([23]).
\end{definicion}
$g_{{\mu}{\nu}}(x,y)$ is the matrix of the fundamental tensor
$g=g_{\mu\nu}dx^{\mu}\otimes dx^{\nu} $. In general, a
non-degenerate structure will be defined by a generalized
fundamental tensor $g$ that is non-degenerate, symmetric and some
smoothness conditions hold.

We should take care of the singularities $F(x,y)=0$. In the above
definition and the principal notions that we develop below eludes
this singularity, following the above definition, based on the
treatment of J. Beem ([21]).

\begin{definicion}
Let $({\bf M}, F)$ be a non-degenerate Finsler structure and
$(x,y,{\bf U}) $ a local coordinate system on ${\bf TM}$. The Cartan
tensor components are defined by the set of functions,

\begin{equation}
{ A}_{\mu \nu \rho }=\frac{1}{2} \frac{\partial g_{\mu \nu }}{\partial y^{\rho}},\quad 
\mu,\nu, \rho =0,...,3.
\end{equation}
\end{definicion}
The components of the Cartan tensor are zero if and only if the Finsler spacetime  $({\bf M},F)$ is a 
Lorentzian structure. It is an homogeneous tensor of order $-1$ in
$y$.

One can introduce several non-linear connection in the manifold ${\bf N}$. First, the non-linear 
connection coefficients are defined by the formula
\begin{equation}
{N^{\mu}_{\nu}}=\gamma^{\mu}_{\nu \rho}y^{\rho}-A^{\mu}_{\nu \rho}
{\gamma}^{\rho}_{rs}y^{r}y^{s},\quad \mu , \nu ,\rho ,r,s=0,...,3.
\end{equation}
The coefficients ${\gamma}^{\mu}_{\nu \rho}$ are defined in local
coordinates by
\begin{displaymath}
 {\gamma}^{\mu }_{\nu \rho}=\frac{1}{2}g^{\mu s}(\frac{\partial g_{s\nu}}{\partial
x^{\rho}}-\frac{\partial g_{\rho \nu}}{\partial
x^{s}}+\frac{\partial g_{s\rho}}{\partial x^{\nu}}),\quad \mu ,\nu
,\rho , s=0,...,3;
\end{displaymath}
$A^{\mu} _{\nu \rho}=g^{\mu l}A_{l\nu \rho}$ and $g^{\mu l}g_{l \nu}=\delta ^{\mu} _{\nu} 
.$ As a consequence, $N^{\mu}_{\nu}$  are not homogeneous
coefficients.

Using these coefficients one obtains a splitting of {\bf TN}. Let us
consider
 the local coordinate system $(x,y) $ of the manifold ${\bf TN}$ and
 an open subset ${\bf U}\in {\bf N}$. The induced tangent basis for ${\bf T}_u {\bf 
N}$ is
\begin{displaymath}
\{ \frac{{\delta}}{{\delta} x^{1}}|_u ,...,\frac{{\delta}}{{\delta} x^{n}} |_u, 
\frac{\partial}{\partial y^{1}} |_u,...,\frac{\partial}{\partial
y^{n}} |_u\},\quad
 \frac{{\delta}}{{\delta} x^{\nu}}|_u =\frac{\partial}{\partial
x^{\nu}}|_u -N^{\mu}_{\nu}\frac{\partial}{\partial y^{\mu}}|_u .
\end{displaymath}
The set of local sections $\{ \frac{{\delta}}{{\delta} x^{1}}|_u 
,...,\frac{{\delta}}{{\delta} x^{n}}|_u,\, u\in {\bf U} \} $
generates the local
horizontal distribution $\mathcal{H}_U $ while $\{ \frac{\partial}{\partial y^{1}}|_u ,..., 
\frac{\partial}{\partial y^{n}}|_u, u\in {\bf U} \}$ the local vertical distribution 
$\mathcal{V}_U$.
The subspaces ${\bf \mathcal{V}}_u $ and ${\bf \mathcal{H}}_u$ are such that the splitting of ${\bf 
T}_u {\bf N}$ holds:
\begin{displaymath}
{\bf T}_u {\bf N}=\mathcal{V}_u \oplus \mathcal{H}_u ,\, \forall
\,\, u\in {\bf U}.
\end{displaymath}
This decomposition is invariant by the right action of ${\bf
GL}(2n,{\bf R})$ and defines a connection in the bundle ${\bf
TN}\rightarrow {\bf M}$. The corresponding dual basis in the dual
vector bundle   ${\bf T^{*}N}$ is
\begin{displaymath}
\{ dx^{0},...,dx^{3}, {{\delta}y^{0}},..., {{\delta}y^{3}}\} , \quad
{{\delta}y^{\mu}}=(dy^{\mu}+N^{\mu}_{\nu}dx^{\nu}).
\end{displaymath}

An extensive and general treatment of the notion of non-linear
connection, allowing for connections in associated bundles that are
$g$-compatible, can be found in the work of Vacaru et al. (for
example [12] and [14]) in the contest of Lagrange spaces and other
generalizations. In this general framework, the coefficients of an
alternative non-linear connection are given by:
\begin{equation}
{N^{\mu}_{\nu}}=\frac{1}{2}\frac{\partial}{\partial y^{\nu}}\big[
g^{\mu \rho} \big(\frac{\partial^2 F^2}{\partial y^{\rho} \partial
y^{\sigma}}y^{\sigma}- \frac{\partial F}{\partial
x^{\rho}}\big)\big],\quad \mu , \nu ,\rho,\sigma =0,...,3.
\end{equation}
These coefficients also define an splitting of ${\bf TN}$ similar to
the described before, such that the formal formulae is maintained.
Note that in contrast with $(3.3)$, this non-linear connection is
homogeneous in $y$ and does not have singularities at $F=0$.

The covariant splitting is equivalent to the existence of a
particular selection of the non-linear connection. These
non-connections, that are also connections in the sense of Ehresmann
([5]), are associated with particular splitting of ${\bf TN}$ due to
associated non-degenerate metric structures in {\bf N}. In our case,
the structure is given by
\begin{equation}
g_{AB}= g_{\mu \nu} dx^\mu \otimes dx^\nu + g_{\mu \nu}\Big( {\delta 
y^{\mu}}\otimes {\delta y^{\nu}}\Big).
\end{equation}
The metric $(3.5)$ is called a Sasaki-type pseudo-Riemannian metric
in ${\bf N}$. A particular definition for the non-linear connection,
with respect the corresponding structure (3.3),
 the horizontal sub-space spanned 
by the distributions $\{\frac{\delta}{\delta x^{\mu}},\, \mu =0,...,3\}$ is orthogonal 
respect the distribution developed by $\{{\frac{\partial}{\partial y^{\mu}}},\, \mu=0,...,3 
\}$.

For the case of timelike trajectories, we can also  introduce the
treatment of Asanov for non-degenerate Finsler spacetimes ([22]),
which we will extend to non-degenerate Lagrange and generalized
Finsler-Lagrange structures. Asanov considered the same notion as in
{\it definition 3.1}, but restricting the smoothness condition of
the hessian $g_{\mu\nu}$ to the vectors $y\in {\bf T}_x{\bf M}$ with
$F(x,y)>0$. In particular, let us denote the sets of admissible
vectors by
\begin{displaymath}
\tilde{{\bf N}}_x:=\{ y\in {\bf T}_x{\bf M}\,|\, F(x,y)>0\},\,\,
\tilde{{\bf N}}=\cup_{x\in {\bf M}}\tilde{{\bf N}}_x .
\end{displaymath}
In the case of non-degenerate Finsler structures, the treatment of
Asanov starts with the definition (3.1) for Finsler spacetimes, but
restricted to the set of admissible vectors $\tilde{{\bf N}}$. Under
these restrictions homogeneous definitions of Cartan, non-linear
connection and Sasaki-type metric can be introduced (for standard
definitions see [4] or [20]). In particular, the Sasaki type metric
is defined by

\begin{equation}
g_{AB}= g_{\mu \nu} dx^\mu \otimes dx^\nu + g_{\mu \nu}\Big( \frac{{\delta 
y^{\mu}}}{F}\otimes \frac{{\delta y^{\nu}}}{F} \Big).
\end{equation}
It defines a non-degenerate metric in ${\bf \tilde{\bf N}}$ and has
associated an arc-length function that is re-parametric invariant
(the main difference with the metric $(3.5)$). However, Asanov's
treatment excludes the null trajectories $F=0$. Although this is not
too expensive for the discussion of this paper, about maximal
acceleration, it is  incomplete as a whole picture.

\section{Covariant Quantum Geometry}

In this section we will follow Beem's convention. Since using
Asanov's treatment, the effective non-degenerate structure in the
spacetime manifold {\bf M}is also non-homogeneous, no advantage is
obtained using this approach to avoid the singularities at $F=0$.
Indeed in this contest, Beem's treatment appears less restrictive
than Asanov's ones.

In order to investigate the properties of the
pseudo-Riemannian structure  $(\tilde{{\bf 
N}},g_{AB})$, we have to note first that the distributions $\{(dx^{\mu},dy^{\mu}),\quad 
\mu=0,...3\}$ do not form a consistent basis
of ${\bf T}^*_u\tilde{{\bf N}}$ for a general non-flat 
manifold. The problem is localized in the set of 1-forms $\{dy^{\mu},\quad \mu =0,...,3\}$. Under 
local coordinate transformations of ${\bf M}$, the induced
transformation rules are
\begin{displaymath}
d\tilde{x}^{\mu}=\frac{\partial \tilde{x}^{\mu}}{\partial
x^{\rho}}dx^{\rho}, \quad
d\tilde{y}^{\mu}=\frac{\partial \tilde{x}^{\mu}}{\partial x^{\nu}}dy^{\nu}+\frac{\partial 
^2 \tilde{x}^{\mu}}{\partial x^{\nu}\partial x^{\rho}} y^{\nu }
dx^{\rho}.
\end{displaymath}
Therefore, the non-covariance problem of the covariance of eq.
$(1.2)$
is at the begin of the Caianiello's construction: 
the distributions  $\{(dx^{\mu},dy^{\mu}),\quad \mu =0,...,3\}$ is not convenient to describe 
differential forms over ${\bf N}$ and produces non-covariant
results.

In order to solve this problem, we propose to consider the analogous construction as in 
Caianiello's model but using the basis $(3.2)$. In a similar way as
in Caianiello's model,
we consider the Sasaki-type 
metric in ${\bf N}$
\begin{displaymath}
dl^2 =g_{\mu \nu } dx^{\mu} dx^{\nu }+(\frac{1}{A_{\max}})^2 g_{\mu \nu}\delta y^{\mu} 
\delta y^{\nu},
\end{displaymath}

It is because the existence of the non-linear connection, represented by the above splitting of 
{\bf TN}, that this construction have invariant meaning. This metric
can be expressed at the point $u\in {\bf N}$ as
\begin{displaymath}
dl^2 |_u =ds^2 +(\frac{1}{A_{\max}})^2 g_{\mu 
\nu}(N^{\mu}_{\rho}N^{\nu}_{\xi}dx^{\rho}dx^{\xi}+dy^{\mu}N^{\nu}_{\rho}dx^{\rho}+
dy^{\nu}N^{\mu}_{\rho}dx^{\rho}).
\end{displaymath}
From this pseudo-Riemannian structure in {\bf N} we obtain a
Lorentzian-type structure
in the 
spacetime manifold {\bf M}. Let 
us recall that $ds^2 _0 =g_{\mu \nu } (x) dx^{\mu }_x dx^{\nu }_x$.
Then,
\begin{displaymath}
ds^2 _0 \Big
(\frac{dy^{\mu}}{ds_0}\frac{dy^{\nu}}{ds_0}\Big)=dy^{\mu}dy^{\nu}.
\end{displaymath}
When the constraint $y^{\mu}=\frac{d x^{\mu}}{ds_0 }=\dot{x}^{\mu}$ is imposed, 
replacing $dx^{\mu}|_u$ by $dx^{\mu}|_x$ and using a particular
timelike congruence to produce the inversion
$x^{\mu}(s_0)\rightarrow s_0 (x)$,
 the metric $(3.3)$ induces an embedding in {\bf 
M} given by
\begin{displaymath}
dl^2 _x =ds^2 _0 \Big( 1+\Big(\frac{1}{A_{max}}\Big)^2 
\ddot{x}^{\sigma}\ddot{x}_{\sigma}\Big)+g_{\mu \nu }(x)\Big(\frac{1 }{A_{max}}\Big)^2 
(N^{\mu}_{\rho}N^{\nu}_{\xi}dx^{\rho}dx^{\xi}+
\end{displaymath}
\begin{displaymath}
+d\dot{x}^{\mu}N^{\nu}_{\rho}dx^{\rho}+d\dot{x}^{\nu
}N^{\mu}_{\rho}dx^{\rho})\Big).
\end{displaymath}
Therefore the new space-time metric can be written as
\begin{equation}
dl^2 _x =ds^2 _0 \Big(1+\Big(\frac{1 }{A_{max}}\Big)^2(   \ddot{x}^{\sigma} 
\ddot{x}_{\sigma}+g_{\mu 
\nu}(N^{\mu}_{\rho}N^{\nu}_{\xi}\dot{x}^{\rho}\dot{x}^{\xi}+\ddot{x}^{\mu}N^{\nu}_{\rho}
\dot{x}^{\rho}+\ddot{x}^{\nu}N^{\mu}_{\rho}\dot{x}^{\rho}))\Big).
\end{equation}
We note that $(4.1)$ is a Lorentzian-type structure, due to the use
of the inversion procedure, except for the fact that it is not
re-parametrization invariant an its dependence on the trajectories.
The above procedure requires to introduce a timelike congruence,
describing the particular evolution of the test particles. This
implies that the Lorentzian-type metric $(4.1)$ depends on the
particular congruence. Although it is not a contradiction, this
dependence is problematic because we do not know a priori the motion
of the test particles and the initial configuration, which
constitutes a rather uncomfortable geometric model and eventually a
probability density function must be introduced.

An alternative treatment to obtain the embedding consists on
consider that $(4.1)$ defines a higher order Lagrange structure (for
a definition of these structures see ref [20]), that being not
homogeneous in $y$ does not define a re-parametrization invariant
arc-length in the tangent space (or via duality, in the phase
space). We interpret the conditions $y^{\mu}=\frac {dx^{\mu}}{ds}$
and $dy^{\mu}=d\dot{x}^{\mu}$ in a more liberal way, in the sense
that they are not directly differential equations associated to the
evolution of a point particle, although we can accommodate any
particular physical configuration in this formalism.

Either considering a Lorentzian-type structure associated with a
congruence or as
 the line element corresponding to a generalized Lagrange
structure, let us denote the quantum contributions by
\begin{displaymath}
h_1 =ds^2 _0 \Big(\frac{1 }{A_{max}}\Big)^2(   \ddot{x}^{\sigma}
\ddot{x}_{\sigma}),\quad
h_2 =ds^2 _0 \Big(\frac{1 }{A_{max}}\Big)^2 g_{\mu 
\nu}(N^{\mu}_{\rho}N^{\nu}_{\xi}\dot{x}^{\rho}\dot{x}^{\xi}+
\end{displaymath}
\begin{displaymath}
+\ddot{x}^{\mu}N^{\nu}_{\rho}\dot{x}^{\rho}+
\ddot{x}^{\nu}N^{\mu}_{\rho}\dot{x}^{\rho}).
\end{displaymath}
Note that our treatment depends on the particular non-linear
connection: different non-linear connections provides different
quantum geometries. Although these contributions can be formally the
same, since they depend on the non-linear connection, different
non-linear connection will promote different competing theories.

If both quantum contributions are comparable, one expects
$|h_1 |\approx \frac{1}{3}|h_2|$. In the 
case of Riemannian structure, the Cartan tensor is zero and the non-linear connection is 
reduced to $N^{\mu} _{\nu} =\gamma ^{\mu} _{\nu \rho}y^{\rho}$.
Then the condition $|h_1|\approx \frac{1}{3}|h_2|$ 
can be expressed as the set of ordinary differential equations
\begin{displaymath}
\ddot{x}^{\mu}+k\gamma^{\mu}_{\nu \rho}\dot{x}^{\nu}\dot{x}^{\rho}=0,\, \mu,\nu,\rho 
=0,...,3.
\end{displaymath}
The factor $k$ is of order 1 when $h_1 \approx -\frac{1}{3} h_2$.
Conversely, if the evolution is classical in the spacetime {\bf M}, both quantum 
corrections are related by $h_1 =-\frac{1}{3}h_2$ and the metric is the 
initial one $g$. In this case $k=1$. We obtain that, for semi-classical particles, the new 
correction $|h_2 | $ is as large as $3|h_1|$, inducing a 
natural almost cancelation of the quantum geometry corrections. Nevertheless, for pure 
quantum particles this is not necessary and a strong difference can
appear between them.

For flat phase-space models ([1]), there is a global 
 coordinate system where the connection coefficients are zero, recovering the original 
Caianiello's flat model:
\begin{displaymath}
dl^2 =ds^2 _0 (1+h_1 +h_2)\longrightarrow dl^2 =ds^2 _0 (1+h_1)
\end{displaymath}
with metric coefficients
\begin{displaymath}
\tilde{g}_{\mu \nu}=(1+ \Big(\frac{1}{A_{max}}\Big)^2 
\ddot{x}^{\sigma}\ddot{x}_{\sigma})\eta _{\mu \nu},
\end{displaymath}
being $\eta_{\mu \nu}$ the Minkowski metric.
This is the typical metric tensor appearing in flat Quantum Geometry. Therefore, the 
predictions and corrections coming from the original flat model are
also maintained in our revision.

On the other hand, from the expression (4.1),  one obtains the
general formula:
\begin{equation}
ds^2=(1+\frac{(D\dot{x})^{\sigma}(D\dot{x})_{\sigma}}{A^2
_{max}})ds^2_0 .
\end{equation}
For Berwald spaces with a linear connection $D$, the geodesic is
just given by
\begin{displaymath}
(D\dot{x})^{\sigma}=\ddot{x}^{\sigma} + N^{\sigma} _{\bf \rho}
\dot{x}^{\rho}=\ddot{x}^{\sigma} + \gamma^{\sigma} _{\nu \rho}
\dot{x}^{\rho}\dot{x}^{\nu}=0.
\end{displaymath}
 This
geodesic interpretation is general for curved Berwald spaces and it
is covariant on the base manifolds.
\begin{teorema}
Consider a Berwald spacetime $({\bf M},F)$ such that the geodesics
of an associated non linear connection are defined by the expression
\begin{displaymath}
\ddot{x}^{\sigma} + N^{\sigma} _{\bf \rho}
\dot{x}^{\rho}=0,\,\,\,\sigma, \rho =0,...,3.
\end{displaymath}
Then,  the particles follow a non-geodesic evolution or the geometry
induced from the Quantum Geometry is the same as the initial
geometry.
\end{teorema}
In the case of a Berwald structure the geodesic equation of the
Berwald connection is just given by equation (4.39) (a Berwald space
is a Finsler structure where one can define a $g$-compatible
connection that lives in the base manifold ([19])). Berwald
structures have the benefice that preserve the Equivalence Principle
and this is the reason why the {\it theorem} is stated in this
category. In the general case the theorem should be
\begin{teorema}
Consider a Finsler spacetime $({\bf M},F)$ and suppose that the
classical evolution of point particles is governed in the spacetime
by the expression
\begin{displaymath}
\ddot{x}^{\sigma} + N^{\sigma} _{\bf \rho} \dot{x}^{\rho}=0,\,\,\,
\sigma, \rho =0,...,3.
\end{displaymath}
Then the effects of the Quantum Geometry on the initial space-time
geometry are trivial.
\end{teorema}

{\it Theorem} 4.1 suggests that a non-trivial Quantum Geometry in
Berwald spaces, where the Equivalence Principle holds, will imply
departures from pure gravity. Unification and conjunction with other
non-geometric interactions, implying deviation from classical
geodesic evolution, are required. This conclusion follows if the
induced structure in the spacetime {\bf M} is of the Berwald type,
because $D_{\dot{x}}\dot{x}=0$ is the geodesic equation and the
Equivalence Principle holds. However, following a classical result
of Einstein et al. ([15)],  in General relativity pointless
particles follow geodesic evolution. Then, non-trivial Caianiello's
models will depart from Einstein's gravity. Nevertheless, these
models can be in principle compatible with the Equivalence Principle
and with deviation from the geodesic motion for the physical
geodesics.

Deviation from geodesic motion naturally occurs and then new field
equations should be contemplated. A possible mathematical framework
was developed in {\it section 2.3} of [12], where generalized
Einstein's field equations for Finsler-Lagrange geometries were
developed. Indeed, in the same reference but in {\it section 3},
particular examples were discussed in diverse general frameworks.
These solutions will be of interest to our problem if they are free
of singularities in the sense that geometric invariants are finite.
This free-singularity condition is equivalent to the existence of
maximal acceleration. Another possible general scheme for maximal
acceleration is Deterministic Finslerian Models ([6]). In these
models, maximal acceleration is contained in the geometric
formulation of some dynamical systems, using the geometry of the
phase space ${\bf T^* TM}$.

{\it Theorem 4.1} also suggests the need of consider the value of
the maximal acceleration in the sense of Brandt ([17]),
\begin{equation}
\frac{(D\dot{x})^{\mu}(D\dot{x})_{\mu}}{A^2 _{max}}<1.
\end{equation}
This definition of proper maximal acceleration is covariant. In
general, the connection used in our approach is not necessarily the
same as in [17]. There is another argument for the inequality
$(4.3)$. If we impose that the quantum corrections leave invariant
the sing of the spacetime interval, it is required the inequality
$(4.3)$ holds. This requirement can be argued from fundamental
principles like locality and causality of the interactions: any
interaction can be so strong to change the interval sing and only
locality properties will be related with the interaction. The
argument was explained in [6] and [7] in the contest of
Deterministic Finslerian Models.

On the other hand, {\it theorem 4.2} implies that if we define the
classical evolution by the equation $D_{\dot{x}} \dot{x}=0$, then
departure from it implies non-trivial quantum effects. If we wish to
interpret this equation as equivalent to have the extremal of the
classical finslerian arc-length functional, the non-linear
connection should be the Cartan non-linear connection $[20]$ or the
Berwald non-linear connection $([19])$ and similar conclusions as
from {\it theorem 4.1} can be obtained, although allowing possible
violations of the equivalence Principle.

\section{Conclusions}
The approach advocated in this note to Caianiello's Quantum Geometry has shown that the non-linear 
connection, in both non-degenerate Finslerian and pseudo-Riemannian
frameworks $({\bf M},g)$,
can play an important role on the 
foundations of the theory and in some of the predictions of the
Quantum Geometry model
 in the case of non-zero curvature.

These deviations from standard Quantum Geometry are not
negligible and are in principle so large as the original 
corrections of Caianiello's model. For instance, semi-classical regimen requires 
small deviations from classical evolution and therefore $h_1 \approx - \frac{1}{3}h_2$. Corrections 
coming from the non-linear connection terms could be also important for the modified 
Schwarzschild's geometry and for the study of neutrinos oscillations in this framework and in 
Kerr spaces([8],[9], [18]). Important changes can also be obtained in some 
astrophysical and cosmological consequences of maximal acceleration
([10],[11]).
However, 
in the case of flat space models, the correction $h_2$ is zero and no new 
corrections have to be added to the original Quantum Geometry.

Covariant Caianiello's Quantum Geometry should be considered as a
phenomenological description with origin in a deeper unified
framework, because the existence of a maximal acceleration is not
natural only with gravitational interaction. However with other
interactions playing the game, maximal acceleration could be in
harmony with the absence of singularities in geometric invariants.

In the case of violation of the Equivalence Principle, the approach
indicated by {\it theorem} 4.2 could be useful. If the particles
follow the classical geodesic $D_{\dot{x}}\dot{x}=0$, then Quantum
Geometry will be trivial. But departure from this geodesic evolution
will indicate the need of consider Caianiello's Quantum Geometry in
a unification scheme. In this contest, Cartan and Berwald non-linear
connections become an important tool for the study of the geometry
associated with these models, linear connections in associated
bundles are less relevant in our construction of Covariant
Caianiello's models. Similar conclusions follows from {\it theorem
4.1} in the case that Equivalence Principle holds.

Finally, we wish to stress that independently of the kind of field
theory behind the ideas expressed in this note, it seems that
Covariant Caianiello`s Model is an example where diverse types of
structures like Finsler, Lagrange and generalized Lagrange as well
as their non-degenerate analogous, appear in a natural way. In this
sense, it is a particular realization of higher order
Finsler-Lagrange geometries. This point is particularly important if
we interpret $(4.1)$ as inducing a higher non-degenerate metric
structure in {\bf M}.

\end{document}